\documentclass[12pt]{article}

\usepackage[margin=1in]{geometry}
\DeclareMathAlphabet{\mathcal}{OMS}{cmsy}{m}{n}

\usepackage{times}
\usepackage{mathptmx}
\usepackage{microtype}

\usepackage{amsmath,amssymb,amsfonts}

\usepackage{algorithm}
\usepackage{algpseudocode}

\usepackage{graphicx}
\usepackage{booktabs}
\usepackage{multirow}
\usepackage[table]{xcolor}
\usepackage{subcaption}

\usepackage[numbers,sort&compress]{natbib}

\usepackage[accsupp]{axessibility}

\definecolor{eccvblue}{RGB}{0,96,172}
\usepackage[breaklinks,colorlinks,linkcolor=eccvblue,citecolor=eccvblue,urlcolor=eccvblue]{hyperref}
\usepackage[capitalize,noabbrev]{cleveref}

\usepackage{orcidlink}

\usepackage{xspace}

\newcommand{\keywords}[1]{\par\noindent\textbf{Keywords: }#1}

\newcommand{\best}[1]{\cellcolor{blue!20}\textbf{#1}}
\newcommand{\second}[1]{\cellcolor{blue!10}\underline{#1}}

\title{\texorpdfstring{$\mathbb{R}^{3}$-RECON}{R3-RECON}: Radiance-Field-Free Active Reconstruction via Renderability}

\author{
Xiaofeng Jin\orcidlink{0009-0006-3659-499X}\quad
Matteo Frosi\orcidlink{0000-0002-5936-7945}\quad
Yiran Guo\orcidlink{0009-0006-7983-8472}\quad 
Matteo Matteucci\orcidlink{0000-0002-8306-6739} \\
\small Politecnico di Milano, Italy \\
\small \texttt{\{xiaofeng.jin, matteo.frosi, yiran.guo, matteo.matteucci\}@polimi.it}
}

\date{}

\begin{document}
\maketitle
\begin{center}
\href{https://github.com/jkff00/r3recon}{Codelink: github.com/jkff00/r3recon}
\end{center}

\begin{abstract}
In active reconstruction, an embodied agent must decide where to look next to efficiently acquire views that support high-quality novel-view rendering. Recent work on active view planning for neural rendering largely derives next-best-view (NBV) criteria by backpropagating through radiance fields or estimating information entropy over 3D Gaussian primitives. While effective, these strategies tightly couple view selection to heavy, representation-specific mechanisms and fail to account for the computational and resource constraints required for lightweight online deployment. In this paper, we revisit active reconstruction from a renderability-centric perspective. We propose $\mathbb{R}^{3}$-RECON, a radiance-fields-free active reconstruction framework that induces an implicit, pose-conditioned renderability field over SE(3) from a lightweight voxel map. Our formulation aggregates per-voxel online observation statistics into a unified scalar renderability score that is cheap to update and can be queried in closed form at arbitrary candidate viewpoints in milliseconds, without requiring gradients or radiance-field training. This renderability field is strongly correlated with image-space reconstruction error, naturally guiding NBV selection.  We further introduce a panoramic extension that estimates omnidirectional (360$^\circ$) view utility to accelerate candidate evaluation. In the standard indoor Replica dataset, $\mathbb{R}^{3}$-RECON achieves more uniform novel-view quality and higher 3D Gaussian splatting (3DGS) reconstruction accuracy than recent active GS baselines with matched view and time budgets.
\end{abstract}

\begin{figure}[tb]
  \centering
  \includegraphics[width=\linewidth]{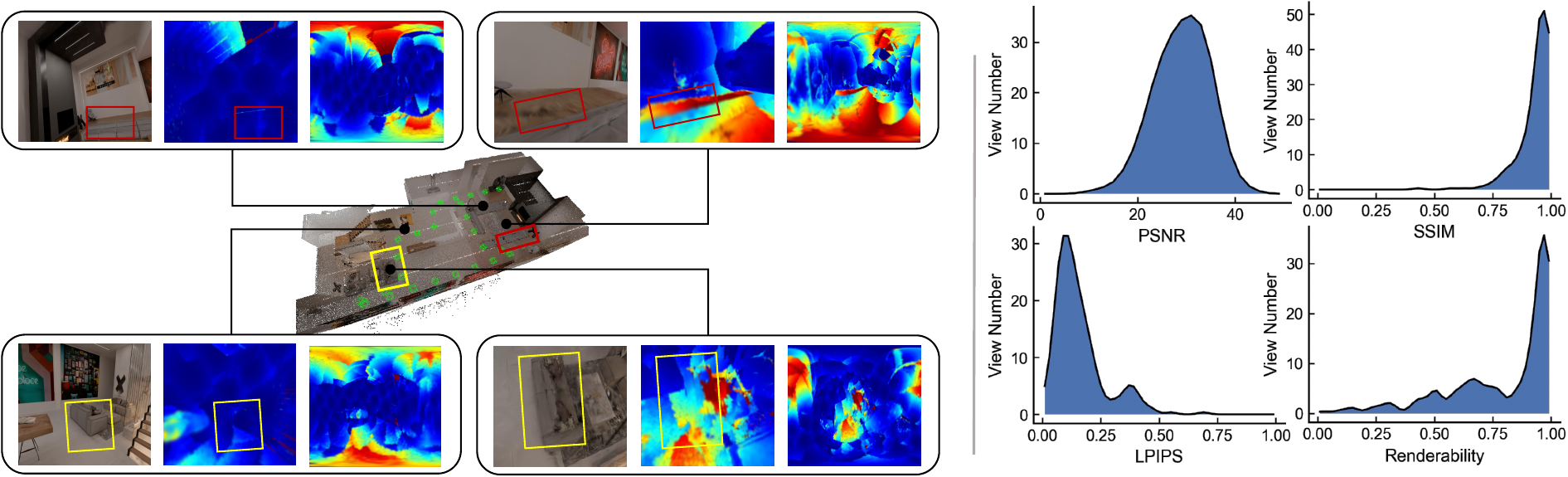}
    \caption{Overview of renderability. Left: green cameras denote training views and black dots denote novel test views; we visualize RGB renderings, per-view renderability, and panoramic renderability. Colored boxes highlight local regions whose novel-view artifacts vary with viewpoint and are well captured by renderability. Right: distributions over 227 novel views of LPIPS, SSIM, PSNR, and our renderability, showing strong alignment with image-space reconstruction quality.}
    
  \label{fig:example}
\end{figure}

\section{Introduction}
\label{sec:intro}

Recent advances in 3D Gaussian splatting (3DGS)~\cite{kerbl20233d} enable high-fidelity novel view synthesis and dense reconstruction at real-time frame rates~\cite{kerbl20233d,compressed}, making Gaussian-based maps attractive for online robotic perception. However, 3DGS quality depends strongly on view density and coverage: sparse or poorly distributed inputs leave regions weakly constrained, leading to artifacts in novel views. Therefore, active view selection during exploration is essential to reduce redundant data while still achieving high-quality reconstruction~\cite{taxonomynbv,gennbv,neuralvisibilityfield}. We study online next-best-view (NBV) selection for 3DGS with an on-board RGB-D camera, where an agent incrementally collects observations and must decide where to look next to maximize downstream rendering quality.

A key challenge in online active reconstruction is to score candidate views under strict latency and memory budgets while providing reliable improvements in reconstruction quality. Most existing methods estimate information gain from the internal state of an online-optimized radiance representation, including NeRF-style fields and 3DGS: they compute uncertainty- or information-theoretic scores from model parameters, gradients, or training losses, and select the next view by maximizing these scores \cite{fisherrf,GAUSSMI,ACTIVEGS}. This model-state dependence is problematic in online settings. Maintaining and repeatedly querying a large radiance representation is expensive on resource-limited platforms, and optimization-state signals can be unreliable before convergence. Recent sensitivity analysis \cite{PUPGS} on 3DGS shows that loss-based second-order signals become stable only after the representation has sufficiently converged, making early-stage view utility estimates noisy and potentially leading to myopic selection.

We decouple view utility from transient optimization state and express information gain in terms of how well view-dependent radiance is constrained by past observations. As \cref{fig:example} shows, we represent this constraint with an implicit renderability field derived directly from lightweight directional observation statistics aggregated online. Motivated by this, we propose $\mathbb{R}^{3}$-RECON, a radiance-fields-free active reconstruction framework that maintains a voxel-statistics map and queries a closed-form renderability score at arbitrary candidate viewpoints in milliseconds. We further introduce a panoramic, pose-level formulation that estimates omnidirectional (360$^\circ$) view utility to accelerate candidate evaluation. On the Replica dataset, the proposed method improves active 3DGS reconstruction over recent baselines under matched view and time budgets.

Our contributions are:
(1) A radiance-fields-free, closed-form view-utility formulation based on observation-statistics-driven renderability, enabling stable online scoring without relying on transient model optimization state.
(2) A lightweight voxel-statistics map that supports millisecond-level NBV evaluation at arbitrary candidate viewpoints.
(3) A 360$^\circ$ panoramic, pose-level utility and extensive Replica experiments demonstrating improved active 3DGS reconstruction under matched budgets.

\section{Related Works}

\subsection{Geometry-driven Active Reconstruction.}
NBV planning is a classic topic in online exploration and active mapping, where a robot incrementally builds a probabilistic occupancy map and selects viewpoints to reduce geometric uncertainty~\cite{placed2022_active_slam_survey}. Representative strategies include frontier-based exploration~\cite{yamauchi1997_frontier} and information-gain objectives that score actions by expected entropy reduction~\cite{stachniss2005_infogain,bourgault2002_ibare}. These ideas have also been extended to active volumetric 3D reconstruction by computing information gain on voxel or octree maps~\cite{isler2016_ig_volumetric,hornung2013_octomap}. While effective for geometric coverage and unknown-space discovery, such criteria are only indirectly tied to photometric fidelity, motivating rendering-aware NBV when neural rendering becomes the reconstruction target.

\subsection{Online Photometric Mapping.}
Neural rendering makes novel-view quality depend on the density and directional distribution of captured views~\cite{nerf}. Recent online reconstruction backbones make it practical to optimize radiance representations incrementally from streaming frames: rather than retraining on all past data, they refine the model by updating a sliding window and/or a small active parameter set, often with keyframe buffering or replay~\cite{imap,niceslam,eslam,coslam,nerfslam}. In parallel, 3DGS provides an explicit primitive-based radiance representation for real-time, high-fidelity rendering~\cite{kerbl20233d}, enabling online 3DGS mapping pipelines such as SplaTAM and MonoGS~\cite{splatam,monogs} as well as further variants for challenging settings~\cite{wildgsslam,gausslam,pings}.

\subsection{Rendering-driven Active Reconstruction.}
Building on online photometric mapping, rendering-driven NBV replaces geometric maps with an online photometric reconstruction state and scores candidate poses by information gain or uncertainty reduction on the evolving model.Accordingly, most rendering-driven NBV methods derive view utility from internal optimization signals of the evolving photometric model. For radiance fields, representative approaches estimate information gain or uncertainty in parameter space, like Fisher information or related criteria~\cite{activenerf,fisherrf,naruto}; 
for 3DGS, utilities are often computed from primitive-level confidence, probabilistic uncertainty, or mutual-information-style objectives~\cite{ACTIVEGS,GAUSSMI,activegamer,active3d}. 
Other pipelines couple view selection with online mapping/planning~\cite{activesplat,agslam,jiang2025_iccv} or use learned ranking signals for fast candidate filtering~\cite{activeviewselector}. 
Despite differences in formulation, a common thread is the reliance on repeatedly maintaining and querying a nontrivial optimization state (parameters/gradients/losses), which is costly under online latency and memory constraints and can be unreliable before convergence; notably, PUP-3DGS~\cite{PUPGS} shows that loss-based second-order sensitivity becomes meaningful mainly after sufficient optimization.

\subsection{Renderability-Centric Active Reconstruction.}
To mitigate the limitations of model-state-driven information gain, we adopt a renderability-centric view, characterizing view utility by how reliably a candidate viewpoint can be rendered given past observations. Yi et al.\ introduced renderability in image-based rendering using a coarse geometric proxy and observation statistics to estimate novel-view synthesis reliability for viewpoint/trajectory selection~\cite{where2render}. Jin et al.\ adapted this idea to 3DGS to guide pseudo-view sampling under sparse inputs~\cite{rfgs}. Existing renderability models, however, are largely grounded in image-based rendering and do not directly characterize uncertainty in radiance representations. Motivated by analyses of uncertainty sources in 3DGS~\cite{klasson_uncertainty}, we reinterpret renderability through an uncertainty lens and extend it to an online formulation enabling efficient view selection, yielding lightweight active photometric reconstruction.

\begin{figure}[!t]
  \centering
  \includegraphics[width=\linewidth]{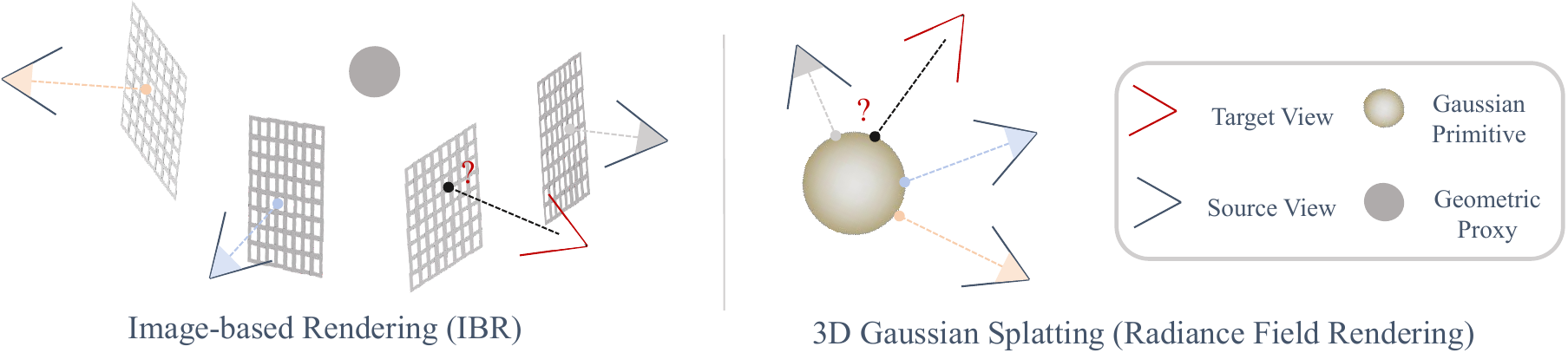}
    \caption{Renderability in IBR and radiance-field rendering. IBR estimates view feasibility from a coarse geometric proxy and source-view overlap, while 3DGS depends on how well primitive appearance is constrained across viewing directions.} 
  \label{fig:compares}
\end{figure}

\section{Technical Approach}
\subsection{Overview}
\label{sec:overview}
We propose an observation-statistics-driven renderability utility for millisecond online NBV scoring without relying on transient reconstruction optimization state. First, we recap the notion of renderability from the image-based rendering (IBR) perspective and contrast it with renderability in view-dependent radiance representations. We then derive an analytic, primitive-level renderability formulation and show how it can be computed online from lightweight observation statistics. Finally, we integrate the resulting renderability score into an active reconstruction pipeline for efficient NBV selection, including a panoramic pose-level evaluation for fast view-direction planning.

\subsection{Preliminaries}
Renderability was introduced by \cite{where2render} as a pre-evaluation metric for IBR. By comparing a target viewpoint against past observations, as illustrated in \cref{fig:compares}, renderability essentially measures how well the source pixels used for synthesis are aligned with the target view. When the scene is purely Lambertian, IBR becomes less sensitive to viewpoint changes, and the dependence on viewpoint differences is reduced.

For radiance representations, the meaning of renderability changes. Radiance fields are fundamentally view-dependent, so sufficient directional observations are still required even in weakly reflective regions. From this perspective, renderability is determined by how well local primitives can fit view-dependent appearance at the query viewing direction, given the history of sampled rays.

\subsection{Renderability in Radiance Representations}
We analyze renderability for a single primitive via its spherical-harmonic (SH) fitting behavior. Following uncertainty sources in 3DGS, we characterize SH fitting quality with two factors: observation bias and noise (\cref{fig:formula}). Observation bias measures the angular discrepancy between the query direction and historical viewing directions; zero bias indicates identical directions and minimal fitting loss. Observation noise summarizes the dispersion of historical radiance observations and reflects material-dependent view variability, which is typically higher for non-Lambertian regions. Accordingly, for each primitive $p_i$, the observation acquired from a viewpoint $s$ (camera pose $T^{(s)}$) is modeled as:

\begin{figure}[!t]
  \centering
  \includegraphics[width=\linewidth]{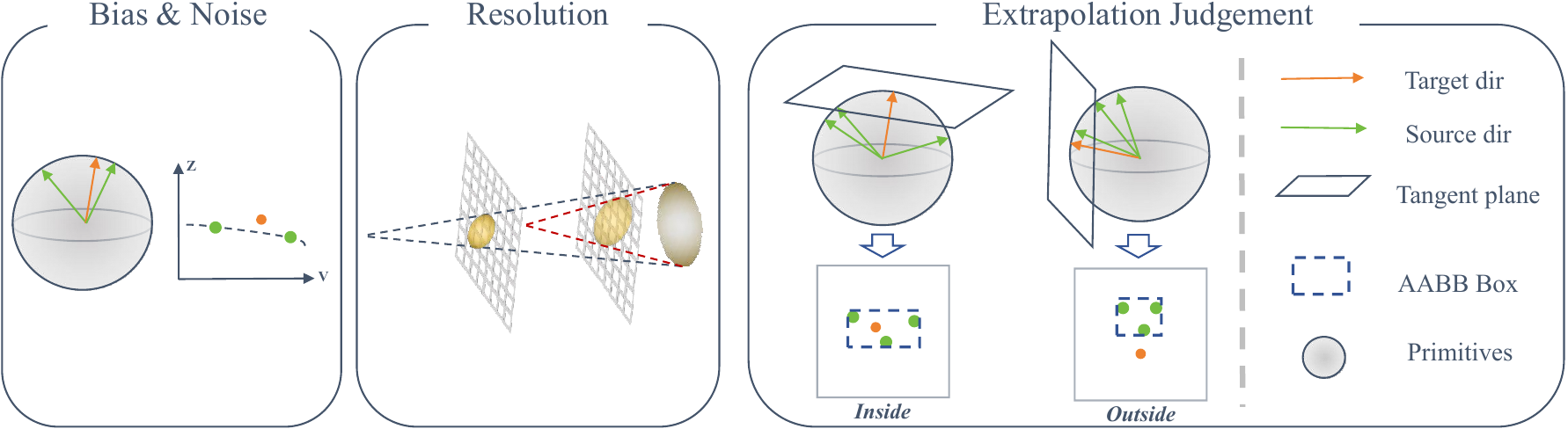}
\caption{Renderability factors. Renderability combines directional bias, appearance noise, and a resolution term. $v$ denotes viewing direction and $z$ the observed radiance; orange points indicate the (unknown) radiance at a novel direction. Noise captures view-dependent variability, and bias measures how far the query direction deviates from the directional support of past observations. Extrapolation is detected by projecting past directions to the query tangent plane and checking whether the query lies outside their support region.}
  \label{fig:formula}
\end{figure}

% \begin{equation}
% \mathbf{z}_i^{(s)}
% =
% \mathbf{z}_i^\star
% \;+\;
% \mathcal{F}\!\left(
% b_i^{(s)}, {\varepsilon}_i^{(s)}
% \right),
% \end{equation}

\begin{equation}
\underbrace{
\mathbf{z}_i^{(s)} - \mathbf{z}_i^\star
}_{\text{rendering\ deviation}}
=
\mathcal{F}\!\left(
b_i^{(s)}, {\varepsilon}_i^{(s)}
\right),
\label{eq:rendering_deviation}
\end{equation}
where $\mathbf{z}_i^\star$ denotes the true (unknown) sample value, $b_i^{(s)}$ represents viewpoint-dependent bias, and ${\varepsilon}_i^{(s)}$ is the noise from all historical observations of $p_i$.

For spherical-harmonic fitting, the same sampling bias can have markedly different effects depending on where the query direction lies relative to the historical samples. From least-squares error propagation and analyses of noise amplification under ill-conditioned sampling\cite{Sparse}, fitting error can grow rapidly when the query falls in an extrapolation regime. To capture this effect, we follow standard spherical geometry that treat the spherical convex hull of historical viewing directions as the valid interpolation domain.

As shown in \cref{fig:formula}, we project historical directions to the tangent plane at the query direction and classify extrapolation if the query falls outside the 2D bounding box of projected samples (a practical local convex-hull approximation).

For bias term, We denote the extrapolation test result by a binary factor $\kappa_i^{(s)}\in\{1,2\}$, where $\kappa_i^{(s)}=1$ for interpolation and $\kappa_i^{(s)}=2$ for extrapolation. let $\{\mathbf{r}_i^{(t)}\}_{t=0}^{n_i}$ denote the unit viewing directions of all past observations, and let $\mathbf{r}_i^{(\mathrm{s})}$ be the unit viewing direction of the current candidate view. Computing the angular deviation between the current and the closest historical direction:

\begin{equation}
\cos{\theta_i^{(\mathrm{s})}}
=
\max_{t \in \{0,\dots,n_i\}}
\langle \mathbf{r}_i^{(t)}, \mathbf{r}_i^{(\mathrm{s})} \rangle
\end{equation}

% \begin{equation}
% b_i^{(s)} =
% \begin{cases}
% \cos\theta_i^{(s)},      & \text{interp.},\\[4pt]
% (\cos\theta_i^{(s)})^2,  & \text{extrap.}.
% \end{cases}
% \end{equation}
\begin{equation}
b_i^{(s)} = \left(\cos\theta_i^{(s)}\right)^{\kappa_i^{(s)}} .
\end{equation}

For the noise term, it is quantified by the mutual differences among its historical observations 
$\{\mathbf{z}_i^{(t)}\}_{t=1}^{n_i}$.  This captures how much the appearance of $p_i$ varies across views. Concretely, the average pairwise discrepancy as
\begin{equation}
\delta_i
= 1 - 
\frac{1}{n_i (n_i - 1)}
\sum_{\substack{t,t'=1\\ t \neq t'}}^{n_i}
\bigl\|
\mathbf{z}_i^{(t)} - \mathbf{z}_i^{(t')}
\bigr\|^2,
\label{eq:pairwise_noise}
\end{equation}
where a smaller $\delta_i$ indicates higher observation noise.

Then convert this discrepancy into a renderability-related term

\begin{equation}
{\varepsilon}_i^{(s)} = \delta_i^{\,\kappa_i^{(s)}\left(1-b_i^{(s)}\right)} .
\end{equation}

This exponential form has two useful properties. First, the base $\delta_i\in[0,1]$ reflects noise consistency: larger $\delta_i$ yields a higher value, meaning more reliable historical observations. Second, the exponent controls the decay rate. When the angular deviation is small ($b_i^{(s)}\!\approx\!1$), the exponent approaches zero and $\delta_i^{\,1-b_i^{(s)}}\!\approx\!1$, indicating almost no rendering loss. Larger deviations increase the exponent and accelerate the decay, penalizing biased viewpoints more strongly.

As illustrated in \cref{fig:formula}, observations are discretized in the pixel domain, so changes in image resolution introduce additional differences in rendering quality. We therefore introduce an auxiliary resolution term to account for range-dependent variations in observation quality. Imaging resolution is inversely proportional to the viewing distance. For a primitive at unbiased depth $D$, the projected pixel footprint satisfies
\begin{equation}
\mathrm{pixel\ size} \propto \frac{1}{D}.
\end{equation}
We define a resolution-related auxiliary term using the depth-based resolution factor
\begin{equation}
\rho_i^{(s)} = \frac{1}{D_i^{(s)}},
\qquad 
\rho_i^{\max} = \max_{t}\frac{1}{D_i^{(t)}},
\end{equation}
which reflects the relative pixel footprint of primitive $p_i$ under different viewpoints. The corresponding resolution gain is then
\begin{equation}
\gamma_i^{(s)} =
\min\!\left(1,\frac{\rho_i^{\max}}{\rho_i^{(s)}}\right)^{\,1-\delta_i b_i^{(s)}} .
\end{equation}

indicating whether the current view demands higher effective resolution than the best historical observation. If $\rho_i^{(s)}\le \rho_i^{\max}$, then $\gamma_i^{(s)}=1$; otherwise, $\gamma_i^{(s)}<1$ indicates lower reconstruction reliability due to insufficient historical resolution.

The exponent $1-\delta_i b_i^{(s)}$ controls the decay rate of this reliability score: primitives with stable appearance (large $\delta_i$) and well-supported viewpoints (large $b_i^{(s)}$) require less additional resolution, yielding weaker attenuation, whereas primitives with high appearance variation or biased viewpoints demand higher resolution and therefore induce stronger decay.

We finally define renderability as the product of three normalized factors: viewpoint bias $b_i^{(s)}$, appearance stability $\varepsilon_i^{(s)}$, and resolution gain $\gamma_i^{(s)}$,
\begin{equation}
R_i^{(s)} \;=\; \mathcal{F}\!\left(b_i^{(s)},\,\varepsilon_i^{(s)}\right)\,\gamma_i^{(s)}
\;=\; b_i^{(s)}\,\varepsilon_i^{(s)}\,\gamma_i^{(s)},
\label{eq:renderability}
\end{equation}
where $b_i^{(s)}, \varepsilon_i^{(s)}, \gamma_i^{(s)} \in [0,1]$. The multiplicative form yields $R_i^{(s)} \in [0,1]$ and assigns high renderability only when the primitive is well-supported directionally, exhibits consistent appearance, and is observed at sufficient resolution.

\begin{algorithm}[!t]
\caption{Online bias term with Fibonacci-binned directions}
\label{alg:bias}
{\small
\begin{algorithmic}
\Statex \textbf{Global (once):} choose $N$ bins by $\theta_{\mathrm{res}}$ and precompute $\{\mathbf q_k\}_{k=1}^N$
\Statex \hspace{1em}$A \leftarrow 2\pi(1-\cos(\theta_{\mathrm{res}}/2));\ \ N \leftarrow \lceil 4\pi/A\rceil;\ \ \varphi_g\leftarrow(1+\sqrt5)/2$
\For{$k=1..N$}
  \State $z_k \leftarrow 1 - 2\frac{k+0.5}{N};\ 
         r_k \leftarrow \sqrt{1-z_k^2};\ 
         \phi_k \leftarrow \frac{2\pi k}{\varphi_g};\ 
         \mathbf q_k \leftarrow (r_k\cos\phi_k,\ r_k\sin\phi_k,\ z_k)$
\EndFor
\Statex \textbf{State (per primitive $p_i$):} visited-bin mask $\mathcal I_i\in\{0,1\}^N$
\Statex \textbf{Update with a new view} $\mathbf r_i^{(t)}$:
$k^\star \leftarrow \arg\max_k \langle \mathbf q_k,\mathbf r_i^{(t)}\rangle;\ \ \mathcal I_i[k^\star]\leftarrow 1.$
\Statex \textbf{Query for direction} $\mathbf r_i^{(s)}$:
$b_i^{(s)} \leftarrow \max_{k:\mathcal I_i[k]=1}\langle \mathbf q_k,\mathbf r_i^{(s)}\rangle;\ \ \Return\ b_i^{(s)}.$
\end{algorithmic}}
\end{algorithm}

\begin{algorithm}[!t]
\caption{Online noise estimation with Welford updates}
\label{alg:noise}
{\small
\begin{algorithmic}
\Statex \textbf{State (per primitive $p_i$):} $n_i,\ \boldsymbol\mu_i\in\mathbb R^3,\ \mathbf M_i\in\mathbb R^{3\times3}$
\hfill (init: $n_i{=}0,\boldsymbol\mu_i{=}\mathbf0,\mathbf M_i{=}\mathbf0$)
\Statex \textbf{Update}$(\mathbf z_i^{(t)}\in\mathbb R^3)$:
\Statex $\ n_i\!\leftarrow\!n_i{+}1;\
\Delta\!\leftarrow\!\mathbf z_i^{(t)}{-}\boldsymbol\mu_i;\
\boldsymbol\mu_i\!\leftarrow\!\boldsymbol\mu_i{+}\Delta/n_i;\ 
\mathbf M_i\!\leftarrow\!\mathbf M_i+(\mathbf z_i^{(t)}{-}\boldsymbol\mu_i)\Delta^\top.$
\Statex \textbf{Query} (assume $n_i\ge 2$):
$\ \mathrm{trs}_i\!\leftarrow\!\operatorname{tr}(\mathbf M_i)/(n_i{-}1);\ 
\delta_i\!\leftarrow\!1-\alpha\sqrt{\mathrm{trs}_i},\quad \alpha=\sqrt{2/1.5}.$
\end{algorithmic}}
\end{algorithm}

\subsection{Online Renderability}
\label{sec:bias}
Computing renderability naively requires aggregating statistics over all past observations. This is impractical online for two reasons: (i) the bias term would have to retain all historical viewing directions on the continuous sphere $\mathbb S^2$, incurring
prohibitive memory and increasing query cost; and (ii) different primitives accumulate
different numbers of observations, which prevents efficient vectorization and further
amplifies computation. The noise term suffers from the same history dependence, so in
incremental exploration the update cost grows quickly with the number of frames. We therefore adopt a principled approximation that makes renderability computation
independent of the number of stored observations.

In \cref{alg:bias}, we discretize the unit sphere with a Fibonacci lattice~\cite{gonzalez2010measurement} $\{\mathbf q_k\}_{k=1}^{N}$. Each primitive maintains a fixed-size visited-bin mask over these uniformly distributed bins, enabling efficient bias tracking and evaluation with constant memory cost and observation count-independent, fully vectorizable computation.

For the noise term, as shown in \cref{alg:noise}, each primitive maintains fixed-size Welford moments to track RGB covariance online. We map the covariance trace to a normalized consistency score $\delta_i\in[0,1]$ using $\alpha$.

\subsection{Active Reconstruction via Renderability}

\begin{figure}[tb]
  \centering
  \includegraphics[width=\linewidth]{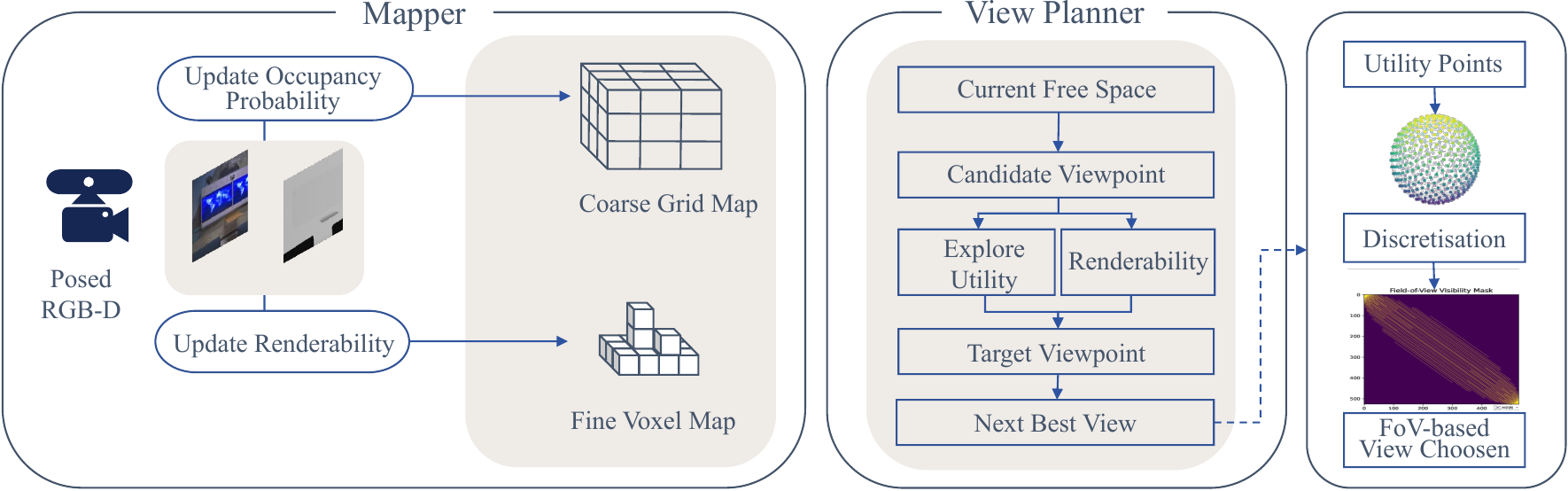}
    \caption{System overview. Posed RGB-D updates a coarse occupancy grid for exploration and a fine voxel-statistics map for renderability. Candidate views are scored by exploration gain and renderability, with panoramic direction selection accelerated by sphere discretization and FoV aggregation.}
  \label{fig:framework}
\end{figure}

We integrate the proposed renderability into online active reconstruction by building on Active-GS\cite{ACTIVEGS} and replacing its view utility with our renderability-based information gain. As shown in \cref{fig:framework}, given a posed RGB-D stream, we maintain two maps. The first is a coarse occupancy map used for exploration and safety, following Active-GS. The second is a lightweight voxel-statistics map used for renderability computation. The voxel-statistics map is downsampled to a resolution of $5$\, cm, and each voxel corresponds to one local primitive that stores the online statistics. To evaluate candidate views efficiently, we extend the rasterizer in ODGS\cite{odgs} for any queried pose, which can quickly compute the visible voxel set and a panoramic depth proxy.

For a candidate viewpoint with position $t$, we compute the visible unexplored grid cells $\mathcal{G}$, and the visible set on the voxel-statistics map for renderability evaluation, $\mathcal{V}$. The exploration utility $U_G=|\mathcal{G}|$ is simply the count of grid cells. While rendering utility as:
\begin{equation}
U_R \;=\; \sum_{v \in \mathcal{V}} \bigl(1 - R_v\bigr).
\label{eq:render_gain_voxel}
\end{equation}
The overall pose score is a weighted combination, as in Active-GS, to prioritize exploration at early stages:
\begin{equation}
U_{\mathrm{view}} \;=\; \lambda_1\,U_G + U_R - \lambda_2\,U_{\mathrm{path}},
\label{eq:view_score}
\end{equation}
where $\lambda_1$ and $\lambda_2$ are weights for the exploration term and the path-cost term in Active-GS, respectively, and $U_{\mathrm{path}}$ is the travel cost to reach the candidate viewpoint computed on the exploration grid.

We further select the viewing direction using the Fibonacci-sphere discretization (Sec.~\ref{sec:bias}). We form a unified 3D point set by combining centers of $\mathcal{G}$ and $\mathcal{V}$, project each point to a unit ray direction on $\mathbb S^2$, and assign it to the nearest Fibonacci bin $k$. For each bin $k$, we precompute the FoV-covered bin set $\mathcal{N}(k)$ when the optical axis aligns with $k$, and compute a direction score by suming per-bin scores over $\mathcal{N}(k)$. The bin with the maximum aggregated score is selected as the final viewing direction.

\section{Experiments}
\subsection{Experimental Setup}

Across all scenes, the Fibonacci sphere used for voxel renderability updates contains $64$ samples. For panoramic direction selection, the Fibonacci discretization uses a target central angle of $10^\circ$. The voxel size is set to $5$\, cm to represent local primitives. All other hyperparameters follow Active-GS.
\begin{figure}[tb]
  \centering
  \includegraphics[width=\linewidth]{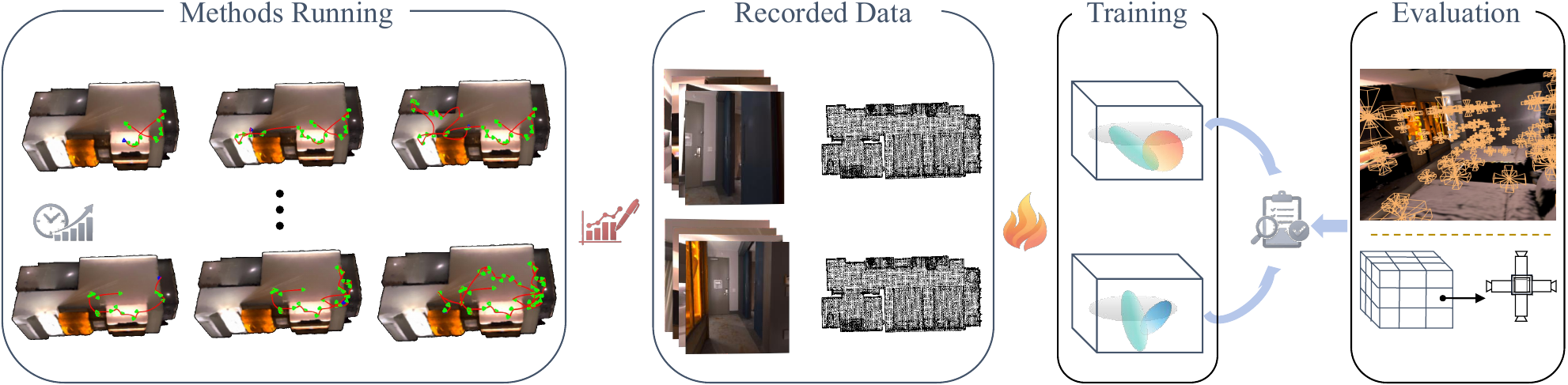}
    \caption{Evaluation pipeline. Each method runs online to generate a trajectory and collect posed RGB and scene pointcloud. A 3DGS model is then trained from the recorded data under matched budgets. For evaluation, test views are sampled from free-space grid locations, each rendered with six canonical viewing directions, to measure scene-level generalization on held-out novel views and reconstruction metrics.} 
    \label{fig:pipe}
\end{figure}

All methods are evaluated in Habitat-Sim\cite{savva2019habitat}. The RGB image resolution is $512\times512$, and the camera FoV is $[60^\circ,60^\circ]$. Experiments are run on a workstation with an NVIDIA RTX A6000 GPU (24\, GB) under Ubuntu 20.04.

Active-GS~\cite{ACTIVEGS} and ActiveGAMER~\cite{activegamer} are used as strong 3DGS-based baselines. Both support full 6-DoF camera motion and achieve competitive performance on Replica. ActiveSplat\cite{activesplat} tailored to ground navigation is not included, as their motion assumptions do not match our free-flying setting. In our tables, Ours replaces the scoring module of Active-GS with our proposed information gain, and Ours-Pano further adds the panoramic extension.

\subsection{Replica-Dense Testset}
The goal of reconstruction is high-fidelity rendering from arbitrary free-space viewpoints. To this end, we construct a dense test set, Replica-Dense, on nine Replica\cite{straub2019replica} indoor scenes. For each scene, collision-free free space is approximated and sampled by a 3D grid with spacing $1.0$\,m$\times 1.0$\,m$\times 0.75$\,m in $(x,y,z)$. At each valid grid location, six test cameras are placed with optical axes aligned to $\pm x$, $\pm y$, and $\pm z$. All methods are evaluated on the same Replica-Dense set.

\begin{table*}[!t]
\centering
\small
\setlength{\tabcolsep}{3.2pt}
\renewcommand{\arraystretch}{1.12}
\resizebox{\textwidth}{!}{%
\begin{tabular}{llccccccccc}
\toprule
Method & Metric & Hotel0 & Off0 & Off1 & Off2 & Off3 & Off4 & Room0 & Room1 & Room2 \\
\midrule
\multirow{3}{*}{ActiveGAMER} & SSIM  & 0.8338 & 0.8859 & 0.8032 & 0.7468 & 0.7339 & 0.7396 & 0.6846 & 0.8008 & 0.7169 \\
                           & PSNR  & 20.51  & 26.16  & 25.74  & 15.95  & 16.45  & 16.02  & 14.86  & 19.75  & 14.55  \\
                           & LPIPS & 0.2738 & 0.1791 & 0.2549 & 0.3658 & 0.3727 & 0.3530 & 0.4369 & 0.3397 & 0.3995 \\
\midrule
\multirow{3}{*}{Active-GS}  & SSIM  & 0.8867 & 0.9138 & 0.8957 & 0.9006 & 0.8792 & 0.8982 & 0.8672 & 0.8778 & 0.9060 \\
                           & PSNR  & 24.60  & 29.87  & 31.39  & 26.91  & 25.38  & 26.45  & 23.01  & 24.51  & 25.07  \\
                           & LPIPS & 0.2134 & 0.1513 & \second{0.1849} & 0.1936 & 0.2209 & 0.2012 & 0.2597 & 0.2508 & 0.2049 \\
\midrule
\multirow{3}{*}{Ours}       & SSIM  & \best{0.9124} & \best{0.9493} & \best{0.9338} & 0.9162 & 0.9140 & \best{0.9269} & \second{0.9053} & \second{0.9149} & \best{0.9426} \\
                           & PSNR  & \best{26.93} & \second{33.95} & \best{37.14} & \second{29.34} & \second{28.93} & \best{31.16} & \second{27.85} & \second{28.85} & \best{31.24} \\
                           & LPIPS & \best{0.1824} & \second{0.1058} & \best{0.1581} & \second{0.1720} & \second{0.1826} & \best{0.1549} & \second{0.2109} & \second{0.2053} & \best{0.1569} \\
\midrule
\multirow{3}{*}{Ours-Pano}  & SSIM  & \second{0.8975} & \second{0.9480} & \second{0.9298} & \best{0.9305} & \best{0.9225} & \second{0.9184} & \best{0.9159} & \best{0.9201} & \second{0.9314} \\
                           & PSNR  & \second{25.95} & \best{34.17} & \second{36.70} & \best{30.74} & \best{30.91} & \second{30.34} & \best{29.41} & \best{29.43} & \second{29.23} \\
                           & LPIPS & \second{0.1992} & \best{0.1017} & \best{0.1581} & \best{0.1553} & \best{0.1668} & \second{0.1590} & \best{0.1964} & \best{0.1963} & \second{0.1711} \\
\bottomrule
\end{tabular}
}
\caption{Per-scene results on Replica-Dense under a fixed time budget. Higher is better for PSNR/SSIM, and lower is better for LPIPS. Best values are highlighted in bold with color, and second-best values are underlined.}
\label{tab:time}
\end{table*}

\begin{figure}[!t]
  \centering

    \includegraphics[width=\linewidth]{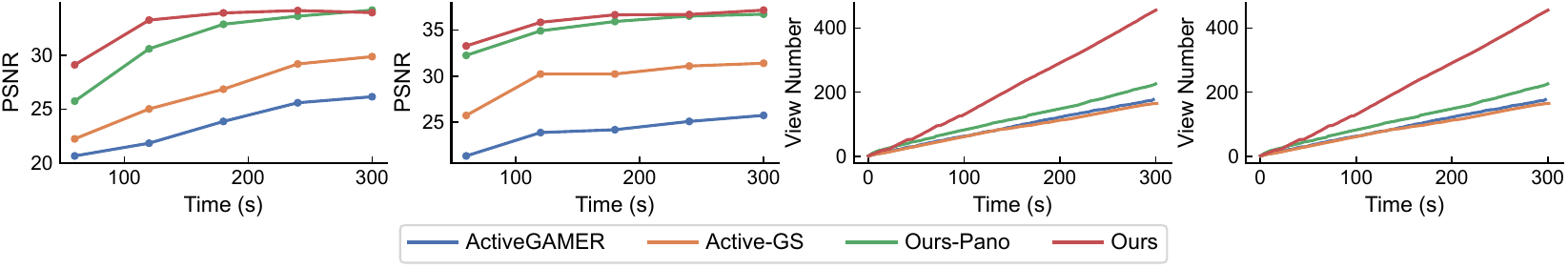}
    
  \vspace{-1mm}
  \caption{Quantitative view quality and exploration efficiency. Each group contains scenes Hotel0 and Office0. The left two pictures is the final novel-view rendering quality versus time. The right two images shows the number of collected images versus time. Under the same time budget, Ours achieves the highest data acquisition efficiency.}
  \label{fig:time_psnr}
\end{figure}

As shown in \cref{fig:pipe}, all methods are treated purely as data collectors. Each method starts from the same initial pose, runs its policy until a fixed budget, and records RGB images, depth-derived point clouds, and camera poses. After data collection, the planner state is discarded, and a vanilla 3DGS model is trained from scratch on the collected views using the same 3DGS implementation and hyperparameters for all methods, without method-specific regularization or augmentation. This decouples planning from reconstruction, so differences on Replica-Dense can be attributed to view selection.

Each method is evaluated under two budget regimes: a time budget of $300$\,s and a view budget of $300$ keyframes. In each regime, we subsample the collected trajectory at $\{60,120,180,240,300\}$ (seconds or views), yielding $2\times5$ acquisition sets per scene. For each set, we train a vanilla 3DGS model and render all Replica-Dense test views. We report PSNR, SSIM\cite{wang2004ssim}, and LPIPS\cite{zhang2018lpips} together with the corresponding acquisition cost, and also record peak GPU memory usage per method and scene to quantify the trade-off between quality and resource consumption.

\begin{table}[!t]
\centering
\small
\setlength{\tabcolsep}{3.2pt}
\renewcommand{\arraystretch}{1.12}
\resizebox{\textwidth}{!}{%
\begin{tabular}{llccccccccc}
\toprule
Method & Metric & Hotel0 & Off0 & Off1 & Off2 & Off3 & Off4 & Room0 & Room1 & Room2 \\
\midrule
\multirow{3}{*}{ActiveGAMER} & SSIM & 0.8831 & 0.8921 & 0.8816 & 0.8735 & 0.8752 & 0.8726 & 0.8796 & 0.8807 & 0.8796 \\
 & PSNR & 23.63 & 26.93 & 31.32 & 22.79 & 24.61 & 25.20 & 24.42 & 24.42 & 23.26 \\
 & LPIPS & 0.2102 & 0.1639 & 0.1851 & 0.2250 & 0.2226 & 0.2122 & 0.2389 & 0.2428 & 0.2313 \\
\midrule
\multirow{3}{*}{Active-GS} & SSIM & \second{0.9085} & 0.9314 & 0.8782 & 0.9142 & \second{0.9220} & 0.9179 & 0.9037 & 0.9139 & 0.9298 \\
 & PSNR & 26.36 & 31.65 & 29.65 & 27.06 & \second{29.57} & 28.92 & 26.68 & 27.64 & 27.40 \\
 & LPIPS & 0.1854 & 0.1244 & 0.1925 & 0.1781 & \second{0.1743} & 0.1680 & 0.2173 & 0.2080 & 0.1752 \\
\midrule
\multirow{3}{*}{Ours} & SSIM & 0.9084 & \best{0.9532} & \best{0.9263} & \second{0.9243} & 0.9054 & \best{0.9282} & \second{0.9200} & \second{0.9227} & \best{0.9357} \\
 & PSNR & \second{26.72} & \best{34.69} & \second{36.01} & \second{29.93} & 27.69 & \best{30.95} & \second{29.60} & \second{29.84} & \second{29.69} \\
 & LPIPS & \second{0.1847} & \best{0.0987} & \best{0.1599} & \second{0.1627} & 0.1934 & \best{0.1549} & \second{0.1948} & \second{0.1946} & \best{0.1636} \\
\midrule
\multirow{3}{*}{Ours-Pano} & SSIM & \best{0.9119} & \second{0.9510} & \second{0.9248} & \best{0.9277} & \best{0.9240} & \second{0.9239} & \best{0.9251} & \best{0.9249} & \second{0.9345} \\
 & PSNR & \best{27.18} & \second{34.34} & \best{36.50} & \best{30.70} & \best{31.26} & \second{30.77} & \best{30.41} & \best{30.39} & \best{29.92} \\
 & LPIPS & \best{0.1825} & \second{0.1021} & \second{0.1603} & \best{0.1596} & \best{0.1685} & \second{0.1553} & \best{0.1877} & \best{0.1913} & \second{0.1654} \\
\bottomrule
\end{tabular}%
}
\caption{Per-scene results on Replica-Dense under a fixed view budget. Higher is better for PSNR/SSIM, and lower is better for LPIPS. Best values are highlighted in bold with color, and second-best values are underlined.}
\label{tab:num}
\end{table}

\begin{figure}[!t]
  \centering
    \includegraphics[width=\linewidth]{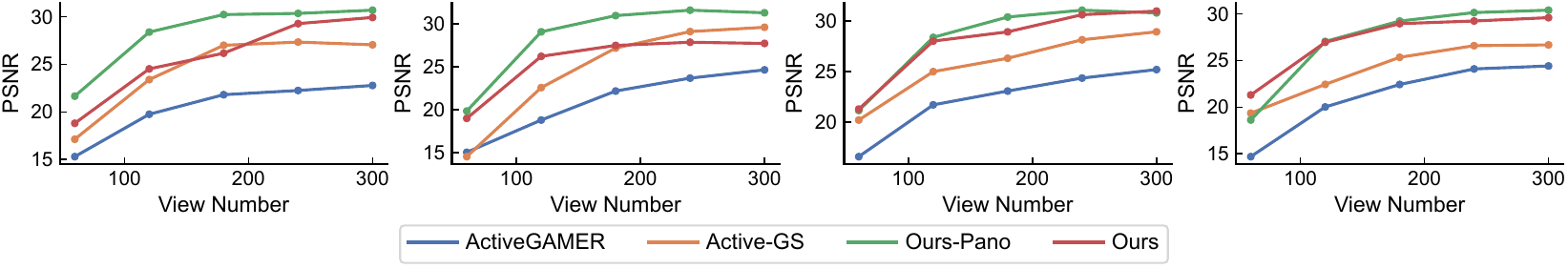}
    \includegraphics[width=\linewidth]{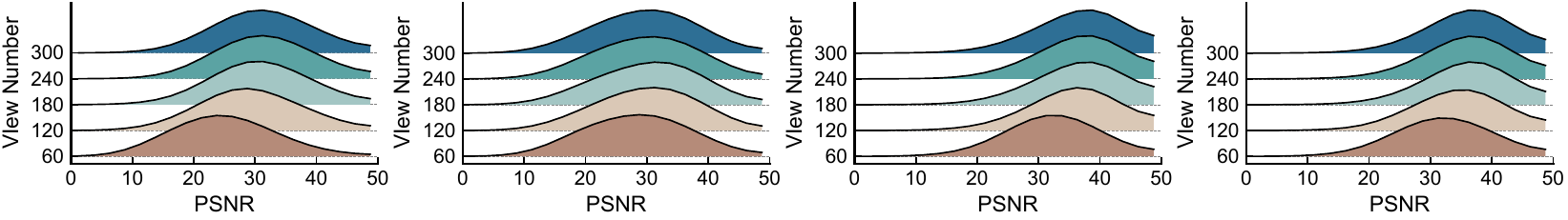}
  \vspace{-1mm}
  \caption{Quantitative results under a view budget. Top row is PSNR versus view budget on four scenes (Office2,3,4 and Room0). Bottom row is PSNR distribution shifts on Office1 across methods (ActiveGAMER, Active-GS, Ours-Pano, and Ours) as the view budget increases.}
  \label{fig:ridge_num}
\end{figure}
\begin{figure}[!t]
  \centering
  \includegraphics[width=\linewidth]{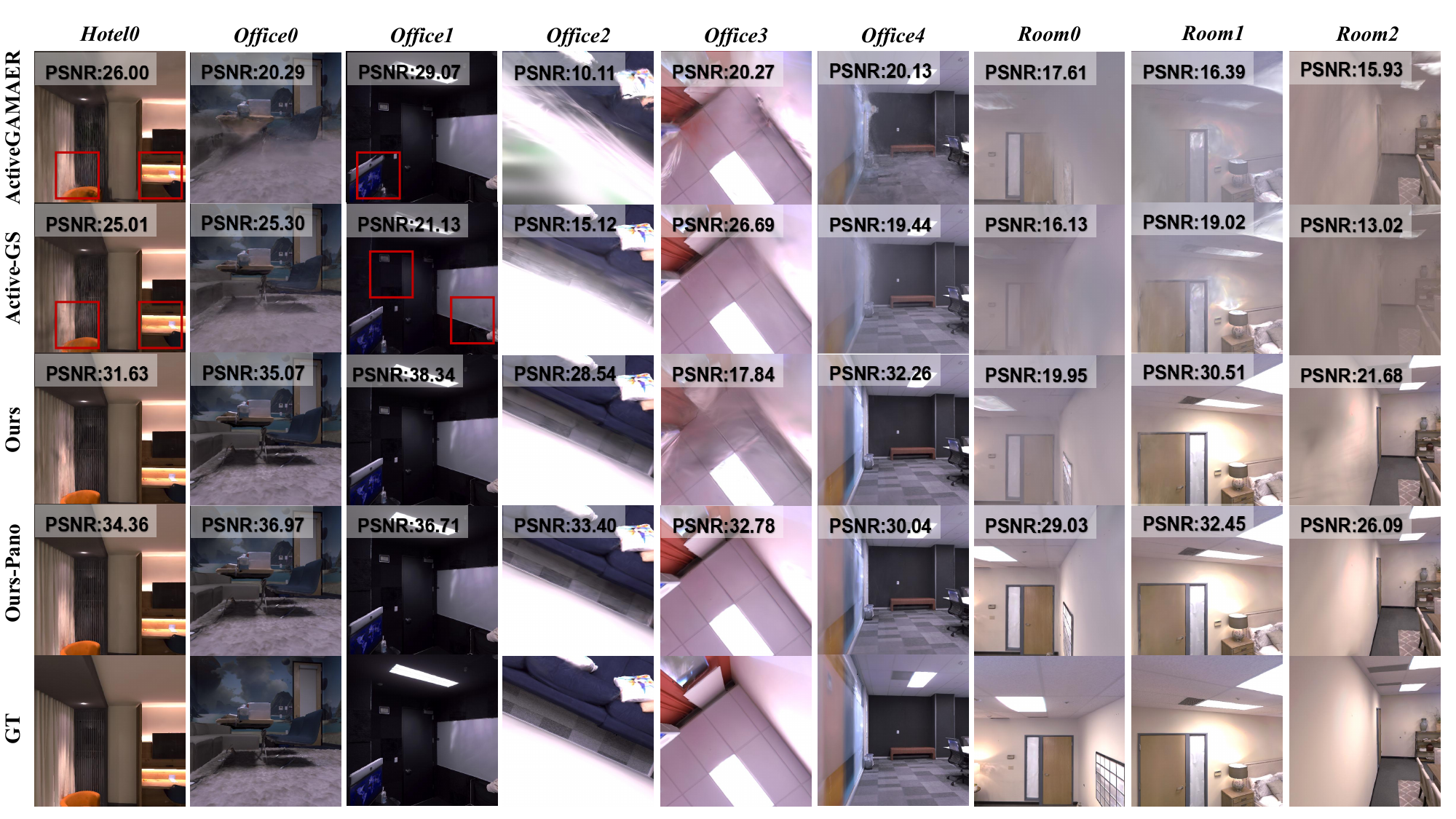}
    \caption{Qualitative comparison under a fixed view budget. The novel view shows that our method produces sharper details and fewer artifacts than prior methods, and the panoramic variant yields more stable reconstruction quality across all scenes. Red boxes highlight regions where differences are subtle.}
    \label{fig:num_image}
\end{figure}

\cref{fig:ridge_num} summarizes performance across budgets: Ours and Ours-Pano remain above both baselines throughout, reflecting higher sample efficiency at low budgets and fewer redundant views at high budgets. The PSNR distribution is also noticeably tighter, with reduced low-quality tails compared to baselines. Qualitative results in \cref{fig:num_image} corroborate these trends, showing sharper textures, cleaner edges, and fewer floaters/ghosting; Ours-Pano further stabilizes quality across scenes.

\subsection{Rendering Quality under Budgets}

With a fixed exploration time of $300$\,s, \cref{tab:time} shows that Ours and Ours-Pano achieve the best or second-best PSNR/SSIM and the lowest LPIPS on nearly all scenes, improving PSNR by about $3$--$5$\,dB over Active-GS (and by a larger margin over ActiveGAMER). \cref{fig:time_psnr} further shows that our PSNR rises faster and saturates higher on Hotel0 and Office0; for the same wall-clock budget, we also collect more keyframes, indicating that renderability-based scoring is both more informative and cheaper to evaluate.

With a fixed budget of 300 keyframes, all methods are trained on the same number of images, so quality depends solely on viewpoint selection. Under this setting, \cref{tab:num} shows that Ours consistently outperforms Active-GS and ActiveGAMER in PSNR/SSIM and LPIPS, validating the effectiveness of renderability-based information gain. Ours-Pano typically improves further, while in small scenes such as Office0 it matches the base version due to saturated coverage.

\begin{figure}[!t]
  \centering
    \includegraphics[width=\linewidth]{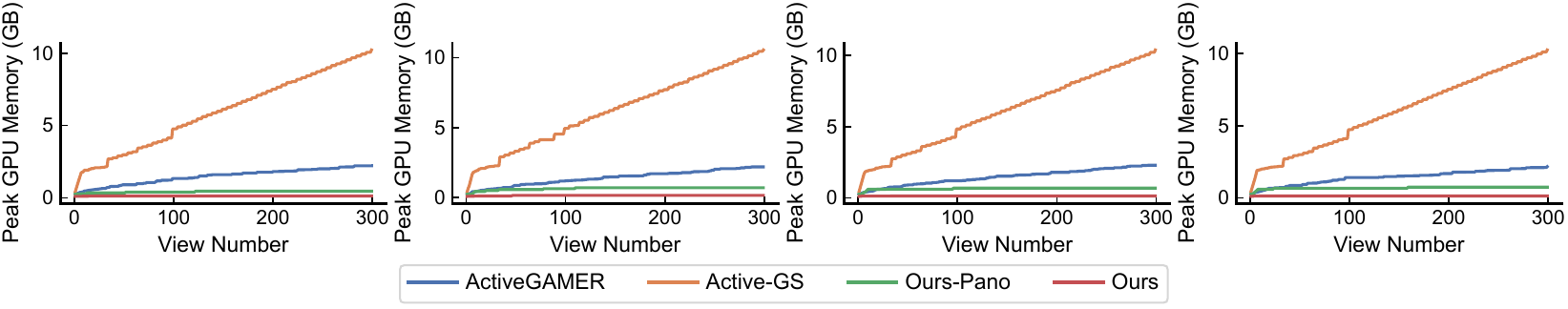}
    \caption{GPU memory usage across four scenes (Office2,3,4, and Room0). Each plot shows the peak GPU memory consumption of different methods as the number of collected views increases.}

  \label{fig:gpu_memory}
\end{figure}

\subsection{Online Performance Analysis}
\cref{fig:gpu_memory} reports peak GPU memory versus the number of views on four scenes. Active-GS shows the steepest growth as it accumulates a heavy radiance field, whereas Ours and Ours-Pano remain below 300\,MB across all scenes since the planner only maintains a lightweight voxel-statistics map.

\begin{figure}[!t]
  \centering
    \includegraphics[width=\linewidth]{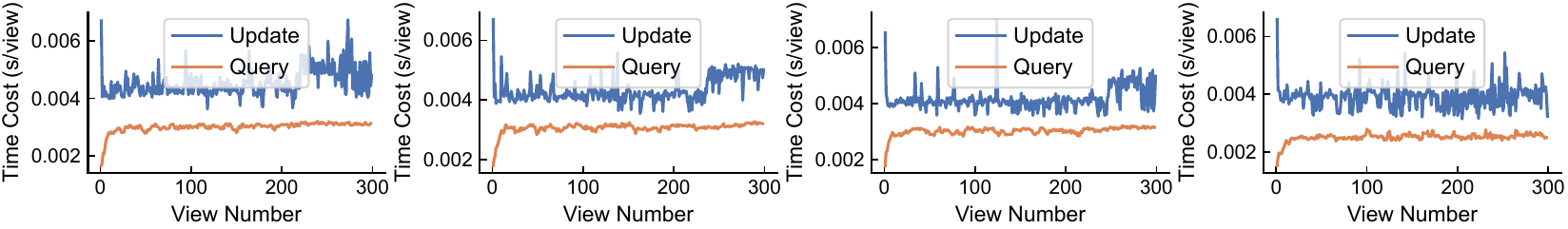}
  \caption{Time cost of Ours-Pano. The figure shows ,in Office2,3,4 and Room0, the change of time cost about renderability updating and querying as the number of keyframe increasing.}
  \label{fig:time-cost}
\end{figure}

\begin{figure}[!t]
  \centering
    \includegraphics[width=\linewidth]{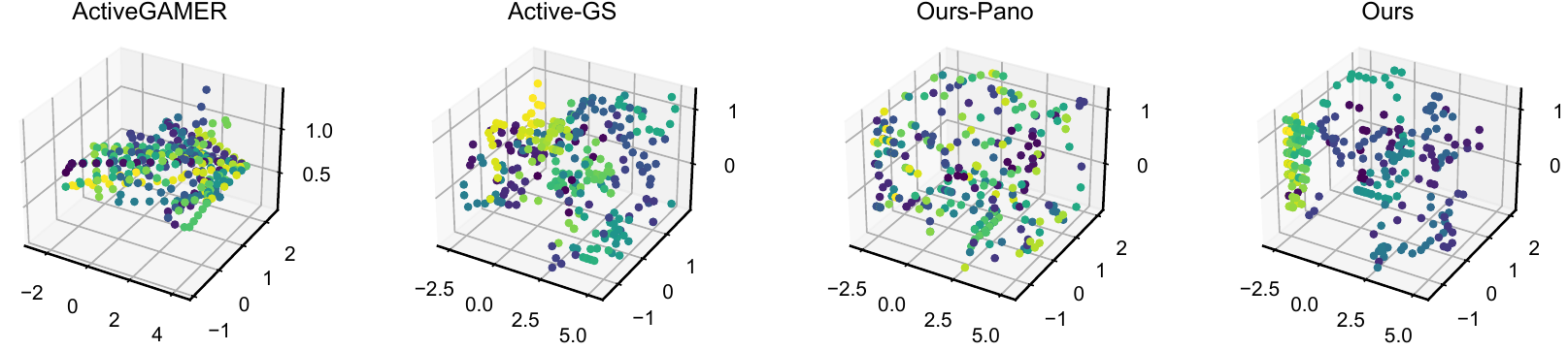}
    \includegraphics[width=\linewidth]{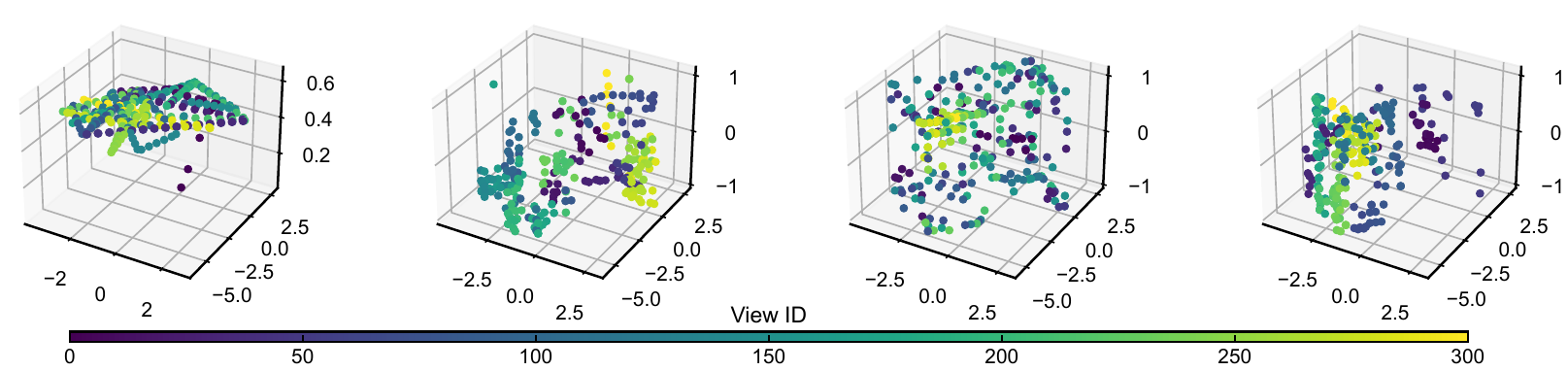}

\caption{Degeneracy analysis of Ours. We visualize 3D heatmaps of camera viewpoints colored by acquisition ID for four methods. The top row is the state in Hotel0, while for Office3, Ours collapses into a local region with repeated, degenerate viewpoints.}

  \label{fig:degeneracy}
\end{figure}

As shown in \cref{fig:time-cost}, per-frame renderability update/query stays in the millisecond range and remains nearly constant as keyframes grow to hundreds, demonstrating observation-count--independent scalability that is critical for long-horizon exploration.
\cref{fig:degeneracy} visualizes camera distributions, comparising with the performance in Hotel0, our non-panoramic variant may degenerate into locally repeated viewpoints due to the trade-off between renderability gain and path cost under random local sampling. By scoring omnidirectional utility at each pose, Ours-Pano mitigates this local trapping and yields more uniform coverage, leading to more stable performance across scenes and budgets.

\section{Conclusion and Limitations}
We presented $\mathbb{R}^{3}$--RECON, a radiance-field-free active reconstruction framework that scores candidate views with an observation-statistics-driven renderability field. By decomposing renderability into directional bias, appearance noise, and resolution, and computing these terms online from a lightweight voxel-statistics map, our method enables closed-form, millisecond-level NBV evaluation without maintaining an online radiance field. A panoramic pose-level extension aggregates omnidirectional utility per pose and accelerates view-direction selection. On the Replica-Dense benchmark, R$^3$-RECON consistently improves 3DGS reconstruction quality over recent active Gaussian-splatting baselines under matched time and view budgets, while using less GPU memory and compute, making it a promising plug-and-play module to pair with existing SLAM systems in real-world autonomous data acquisition. 

A current limitation is that candidate viewpoints are sampled locally at random, which can be suboptimal in large or cluttered environments; future work will couple renderability-based scoring with learned or longer-horizon planning strategies to generate more informative candidates.

\clearpage  % TODO REVIEW/FINAL: This \clearpage needs to be removed from both review and camera-ready versions.

\bibliographystyle{unsrtnat}
\bibliography{main}
\end{document}